\documentclass[aps,prd,showpacs,preprint,nofootinbib]{revtex4-1}
\usepackage{amsmath,amsthm,amssymb,verbatim,graphicx,color,soul,fullpage}
\usepackage{tikz}
\usepackage[compat=1.0.0]{tikz-feynman}
\allowdisplaybreaks

\usepackage[colorlinks]{hyperref}

\long\def\symbolfootnote[#1]#2{\begingroup%
\def\thefootnote{\fnsymbol{footnote}}\footnote[#1]{#2}\endgroup}

\definecolor{LRed}{rgb}{1,.8,.8}
\definecolor{MRed}{rgb}{1,.6,.6}
\definecolor{LBlue}{rgb}{.8,.8,1}
\definecolor{MBlue}{rgb}{.6,.6,1}
\definecolor{LGreen}{rgb}{.8,1,.8}
\definecolor{MGreen}{rgb}{.6,1,.6}

\newcommand{\SU}{\mathrm{SU}}
\newcommand{\U}{\mathrm{U}}
\newcommand{\Tr}{\mathrm{Tr}\,}

\newcommand{\vev}[1]{\langle #1 \rangle}

\begin{document}
\centerline{\large\bf  
Domain walls and CP violation with left right supersymmetry :
}

\vskip 0.1cm

\centerline{\large\bf  
 implications for leptogenesis and electron EDM
}

\vskip 1.0 cm

\centerline{
\bf Piyali Banerjee\symbolfootnote[1]{
banerjee.piyali3@gmail.com
}
and
Urjit A. Yajnik\symbolfootnote[2]{
yajnik@iitb.ac.in
}
}
\medskip
\centerline{\it Department of Physics, Indian Institute of 
Technology Bombay, Mumbai 400076, India}
\bigskip
\bigskip

\begin{center}
{\large \bf Abstract}
\end{center}

\bigskip

Low scale leptogenesis scenarios are difficult to verify due to our inability to
relate the parameters involved in the early universe processes with the low energy
or collider observables. Here we show that one can in principle relate the parameters
giving rise to the transient $CP$ violating phase involved in leptogenesis with
those that can be deduced from the observation of electric dipole moment (EDM)
of the electron. We work out the details of this in the context of the left right symmetric 
 supersymmetric  model (LRSUSY) which provides a strong connection between 
 such parameters.
In particular, we show that baryon asymmetry requirements imply the scale $M_{B-L}$ of $U(1)_{B-L}$
symmetry breaking to be larger than $10^{4.5}~\mathrm{GeV}$.
Moreover the scale $M_R$ of $\SU(2)_R$ symmetry breaking is tightly
constrained to lie in a narrow band significantly below $M_{B-L}^2 / M_{EW}$.
These are the most stringent constraints on the parameter space of
LRSUSY model being considered.

\bigskip

PACS Numbers: 
\vfill

\section{Introduction}
One of the three necessary Sakharov conditions for  
dynamical generation of 
baryon asymmetry of the universe is CP violation \cite{Sakharov}. The Standard Model (SM) 
of particle physics has the ingredients to satisfy all the  three Sakharov conditions and
could, in principle, generate some baryon asymmetry\cite{Kuzmin}. 
However the requirement of a first order Electroweak Phase Transition (EWPT) 
requires the mass of the Higgs to remain less than about 80 GeV 
\cite{Kajantie1,Kajantie2,Csikor} which is far 
below the mass of the recently discovered SM Higgs boson 
\cite{Exptmass1,Exptmass2}. Furthermore the 
CP violation in the CKM matrix of SM is too small to produce a reasonably 
large baryon asymmetry \cite{Gavela}. 

The left right symmetric model of Mohapatra and Senjanovi\'{c}
\cite{MohapatraLR} (LRSM) is a minimal extension of the Standard Model
based on the gauge group 
$\SU(3)_c \times \SU(2)_L \times \SU(2)_R \times \U(1)_{B-L}$ augmented
with the discrete $Z_2$ left-right symmetry. The
usual $\SU(2)_L$ Higgs doublet of SM is extended to a 
$\SU(2)_L \times \SU(2)_R$ bidoublet. The model
naturally accommodates the parity violation of SM as a result of
spontaneous symmetry breaking of $\SU(2)_R \times \U(1)_{B-L}$ and
also elegantly explains small neutrino masses via the seesaw mechanism.
Both goals are achieved by adding new heavy Higgs $\SU(2)$ triplets in 
the theory.

The supersymmetric extension of LRSM, which we call the Simplified
Left Right Symmetric Supersymmetric Standard Model (SLRSUSY),
marries the desirable features of both LRSM and supersymmetry.
However it suffers from the drawback that spontaneous parity violation
cannot take place without violating R-parity also \cite{Kuchimanchi}.
To remedy this, one can take 
the supersymmetry breaking scale $M_S$ to be larger than the
$\SU(2)_R \times \U(1)_{B-L}$ breaking scale $M_{B-L}$ but that would
lose out on several desirable features of supersymmetry.
Alternatively, one can introduce additional heavy Higgs $\SU(2)$
triplets with appropriate gauge group charges. This was done in the two 
papers
\cite{Aulakh:1997ba, Aulakh:1997fq}, which we shall call the 
Left Right Symmetric Supersymmetric Standard Model (LRSUSY). 
The LRSUSY model contains two distinct high mass scales, $M_R$
the scale of $\SU(2)_R$ symmetry breaking, and
$M_{B-L}$ the scale of $U(1)_{B-L}$ symmetry breaking satisfying 
$M_R \gg M_{B-L}$, an intermediate SUSY breaking scale $M_S$ and a low mass
scale $M_H$ where the bidoublet Higgs get vevs. 
Thus, the LRSUSY model has several
desirable features not possessed by SLRSUSY like spontaneous parity 
violation without R-parity
violation, low scale supersymmetry with exact R-parity and
existence of a stable lightest supersymmetric particle.

Breaking of the $Z_2$ discrete symmetry of left-right symmetric models  in the early universe ensures the occurrence of
domain walls, but also begs for the $Z_2$ symmetry to be not exact to avoid conflict with the
observed homogeneous universe. However an approximate $Z_2$ is adequate to
generate the domain walls which are a robust topological prediction independent of details
of parameters, and can play the same role as the phase transition bubble walls for the
purpose of leptogenesis.

In the context of non-supersymmetric LRSM, fate of the $CP$ violating phase
was studied in \cite{ClineYajnik} for the purpose of explaining leptogenesis.
It was shown that a spatially varying CP violating phase occurs inside 
the domain walls separating the left handed and right handed 
domains.
They showed that in order to explain the observed baryon asymmetry of
the universe, the Yukawa coupling of the right handed
neutrino to the $\SU(2)$-triplet Higgs must be larger than $10^{-2}$.
They also obtained some heuristic constraints on the mass of the
left handed neutrinos or alternatively, the temperature scale of 
LR symmetry breaking.
Since the LRSUSY model has additional Higgs bosons as well as additional
CP violating phases, it may be able to generate the necessary 
conditions for successful baryogenesis in the early universe. 
The first step in this direction was taken in \cite{Anjishnu} 
by showing the possibility of having a spatially varying CP violating phase
in the domain walls  in the context of LRSUSY.
However they did not provide any quantitative estimates and left
open the question of whether LRSUSY is actually capable of generating 
the baryon asymmetry of the universe.

The core idea of such proposals is that the spatially varying phase
implies a spatially varying complex mass for left handed neutrinos 
inside the wall. 
Studying the diffusion equation for lepton number density with 
spatially varying complex mass
results in the preferential transmission
of left handed neutrinos across a slowly moving thick domain wall. 
The moving wall encroaches upon the
energetically disfavoured right handed domain.
We solve the diffusion equation numerically for various wall speeds 
and thicknesses,
obtaining excesses of almost massless left handed neutrinos inside the
left handed domain. After the wall disappears, electroweak sphaelerons
convert a part of the neutrino excess to baryon excess. We calculate
the amount of baryon excess that survives the washout processes.
Requiring the surviving baryon excess to be not much above the 
experimental limit of around $6 \times 10^{-10}$ 
for the
baryon asymmetry to entropy ratio \cite{Canetti} allows us to constrain
the $(M_R, M_{B-L})$ parameter space of LRSUSY.

A smoking gun signature of CP violation in a theory is the presence of
a non-zero EDM of the electron and neutron. 
The Standard Model predicts a non-zero electron EDM at the three loop
level but the effect
is estimated to be very small, around $1.9 \times 10^{-39}~\mathrm{e~cm}$
\cite{YamaguchiYamanaka}. This is way below the current
experimental upper bound of $1.1 \times 10^{-29}~\mathrm{e~cm}$ obtained
by the ACME~II experiment~\cite{ACME2}.
Left right symmetric theories contain many additional sources of 
CP violation
as compared to the Standard Model, and so predict larger electron and 
neutron EDMs. 
Thus the experimental bound on EDMs serve to constrain the parameters of 
these theories.
Electron and neutron EDMs in the non-supersymmetric LRSM have been 
studied in several earlier works e.g. \cite{Nieves, Xu}, obtaining 
lower bounds on the scale  
$M_{B-L}$ of $\SU(2)_R \times U(1)_{B-L}$ symmetry breaking and on 
the mass $M_{H'}$ of the 
heavier scalar Higgs in the
two Higgs doublets arising from breaking of left right symmetry 
in the bidoublet Higgs of LRSM. 
Electron and neutron EDMs have been studied in the SLRSUSY model
by \cite{FrankE, FrankN},
obtaining bounds on the masses of certain superpartners and some other
parameters of SLRSUSY.

The LRSUSY model\cite{Aulakh:1997fq} contains essentially only one major unknown, the
$M_R$ scale and a trilinear Higgs coupling parameter $\alpha$ that 
is used to ensure that 
only two out of four SM type
Higgs doublets arising after $\SU(2)_R \times \U(1)_{B-L}$ symmetry
breaking remain lighter, with a relative phase between their vacuum expectation values. 
In this paper we study  the CP violating phase of the 
bidoublet fields in LRSUSY by setting up the domain wall solutions.
The parameters relevant for the domain wall solutions are understood to involve 
the high temperature corrections needed in the early Universe, at the temperature $T_{B-L} \sim M_{B-L}$.
We separately investigate the contribution of the lighter two mass eigenstates of the
bidoublets to the electron EDM at zero temperature, which differ only by the $T_{B-L}$ corrections.
We compute the contribution of this low energy phase to the EDM at one 
loop and two loop levels as a function of $\alpha$. The two loop computation follows along 
the lines of the seminal work of Barr and Zee \cite{BarrZee} on the electron EDM in multi
Higgs doublet models. 
A similar calculation of electron EDM arising as a residual effect of
domain wall collapse in two Higgs doublet models was done in
\cite{Chen}.
It turns out that successful leptogenesis in LRSUSY requires 
$\alpha\gtrsim 0.1$.
Combining this with the requirement that the EDM obtained be less than 
the experimental limit of $1.1 \times 10^{-29}~\mbox{e cm}$, we get an 
allowed region in the $(M_{B-L}, M_R)$-parameter space of LRSUSY.

It turns out that the limit on the parameter space arising from
baryon asymmetry is more stringent than the limit from electron EDM.
Further requiring that the observed baryon asymmetry be explained to
within an order of magnitude by LRSUSY puts stringent
constraints on the $(M_{B-L}, M_R)$-parameter space of LRSUSY. 
In particular $M_{B-L} < 10^{4.5}~\mathrm{GeV}$ is ruled out.
These are the most
stringent constraints on the parameter space of LRSUSY by far.

\section{The LRSUSY model revisited}
The gauge group of LRSUSY is the left-right symmetric group
$\SU(3)_c \times \SU(2)_L \times \SU(2)_R \times \U(1)_{B-L}$.
The Higgs sector consists of two Higgs bidoublets $\Phi_1$ and $\Phi_2$, 
three left handed Higgs triplets $\Delta$, $\bar{\Delta}$ and $\Omega$, 
and three right handed Higgs triplets $\Delta_c$, $\bar{\Delta}_c$ and 
$\Omega_c$.
Following Kuchimanchi-Mohapatra \cite{Kuchimanchi}, the gauge group 
charges are
\begin{equation}
\begin{array}{c c l c c l c c l}
\Phi_1         & = & (1, 2, 2^*,  0), &
\Phi_2         & = & (1, 2, 2^*,  0), & \\
\Omega         & = & (1, 3, 1,  0), &
\Delta         & = & (1, 3, 1,  2), &
\bar{\Delta}   & = & (1, 3, 1, -2), \\
\Omega_c       & = & (1, 1, 3^*, 0), &
\Delta_c       & = & (1, 1, 3^*, -2), &
\bar{\Delta}_c & = & (1, 1, 3^*, 2).
\end{array}
\end{equation}
In other words, the group action is given by
\begin{equation}
\{\Phi_1, \Phi_2\} 
\rightarrow U_L \{\Phi_1, \Phi_2\} U_R^\dagger, 
~~~
\{\Delta, \bar{\Delta}, \Omega\}
\rightarrow U_L \{\Delta, \bar{\Delta}, \Omega\} U_L^\dagger,
~~~
\{\Delta_c, \bar{\Delta}_c, \Omega_c\}
\rightarrow U_R^* \{\Delta_c, \bar{\Delta}_c, \Omega_c\} U_R^T.
\end{equation}

The electric charge is given by
$
Q = T_{3L} + T_{3R} + \frac{B-L}{2}.
$
From Aulakh et al. \cite{Aulakh:1997fq} the Higgs part of the superpotential is 
given by
\begin{equation}
\label{eq:WHiggs}
\begin{array}{rcl}
W 
& = &
m_\Delta (\Tr \Delta \bar{\Delta} + \Delta_c \bar{\Delta}_c) +
\frac{m_\Omega}{2} (\Tr \Omega^2 + \Tr \Omega_c^2) +
\mu_{ij} \Tr \tau_2 \Phi_i^T \tau_2 \Phi_j \\
&   &
{} +
a (\Tr \Delta \Omega \bar{\Delta} + \Tr \Delta_c \Omega_c \bar{\Delta}_c)+
\alpha_{ij} (
\Tr \Omega \Phi_i \tau_2 \Phi_j^T \tau_2 +
\Tr \Omega_c \Phi_i^T \tau_2 \Phi_j \tau_2 
),
\end{array}
\end{equation}
where $\alpha_{ij}$, $\mu_{ij}$ are complex numbers satisfying
$\mu_{12} = \mu_{21}$, $\alpha_{ij} = -\alpha_{ji}$. Define 
$\alpha = \alpha_{12}$. Then $\alpha_{21} = -\alpha$ and 
$\alpha_{11} = \alpha_{22} = 0$. 

The charge zero condition forces the vevs of the Higgs fields to be
\begin{equation}
\label{eq:HiggsVevs}
\begin{array}{c c l c c l c c l} 
\vev{\Phi_1} & = & 
\left(
\begin{array}{c c}
k_1 &  0    \\
0   &  k_1' 
\end{array}
\right), &
\vev{\Phi_2} & = & 
\left(
\begin{array}{c c}
k_2 &  0    \\
0   &  k_2' 
\end{array}
\right), & \\
\vev{\Omega} & = & 
\left(
\begin{array}{c c}
\omega &  0    \\
0      & -\omega 
\end{array}
\right), &
\vev{\Delta} & = & 
\left(
\begin{array}{c c}
0 & 0 \\
d & 0 
\end{array}
\right), &
\vev{\bar{\Delta}} & = & 
\left(
\begin{array}{c c}
0 & \bar{d}    \\
0 & 0 
\end{array}
\right), \\
\vev{\Omega_c} & = & 
\left(
\begin{array}{c c}
\omega_c &  0    \\
0        & -\omega_c 
\end{array}
\right), &
\vev{\Delta_c} & = & 
\left(
\begin{array}{c c}
0   & 0    \\
d_c & 0 
\end{array}
\right), &
\vev{\bar{\Delta}_c} & = & 
\left(
\begin{array}{c c}
0 & \bar{d}_c \\
0 & 0 
\end{array}
\right),
\end{array}
\end{equation}
where the quantities above are in general complex numbers.

The D-terms are given by (where $m = 1, 2, 3$ refer to the three generators
of $\SU(2)$) \cite{Kuchimanchi} 
\begin{equation}
\label{eq:DTerm}
\begin{array}{rcl}
D_{L, m}
& = &
\Tr (
2 \Omega^\dag \tau_m \Omega +
2 \Delta^\dag \tau_m \Delta + 
2 \bar{\Delta}^\dag  \tau_m \bar{\Delta} +
\Phi_1^\dag \tau_m \Phi_1 +
\Phi_2^\dag \tau_m \Phi_2
) \\
D_{R, m}
& = &
\Tr (
2 \Omega_c^\dag \tau_m \Omega_c +
2 \Delta_c^\dag \tau_m \Delta_c + 
2 \bar{\Delta}_c^\dag  \tau_m \bar{\Delta}_c +
\Phi_1 \tau_m^T \Phi_1^\dag +
\Phi_2 \tau_m^T \Phi_2^\dag
) \\
D_{B-L}
& = &
\Tr (
2 \Delta^\dag \Delta - 
2 \bar{\Delta}^\dag \bar{\Delta} - 
2 \Delta_c^\dag \Delta_c + 
2 \bar{\Delta}_c^\dag \bar{\Delta}_c 
).
\end{array}
\end{equation}

After substituting the vevs into the D-term expressions we get
\begin{equation}
\label{eq:DTermVevs}
\begin{array}{rcl}
\vev{D_{B-L}}
& = & 
2(|d|^2 - |\bar{d}|^2 - |d_c|^2 + |\bar{d}_c|^2), \\
\vev{D_{L,1}}
& = & 
\vev{D_{L,2}} = \vev{D_{R,1}} = \vev{D_{R,2}} = 0, \\
\vev{D_{L,3}}
& = &
2(-|d|^2 + |\bar{d}|^2) + |k_1|^2 - |k_1'|^2 + |k_2|^2 - |k_2'|^2, \\ 
\vev{D_{R,3}}
& = &
2(-|d_c|^2 + |\bar{d}_c|^2) + |k_1|^2 - |k_1'|^2 + |k_2|^2 - |k_2'|^2. 
\end{array}
\end{equation}
Taking $|d| = |\bar{d}|$ and 
$|d_c| = |\bar{d}_c|$ ensures the vanishing of $D_{B-L}$-term always. 
We can use the $B-L$ gauge invariance to ensure that $d$, $\bar{d}$ 
have the
same complex phase. Subsequently using $\SU(2)_L$ invariance, we can
ensure that $d = \bar{d}$ and real positive.

Extending Aulakh et al, we see that the resulting F-terms are:
\begin{equation}
\label{eq:FTerm}
\begin{array}{rcl}
F_{\bar{\Delta}} & = &
m_\Delta \Delta + 
a (\Delta \Omega - \frac{\Tr \Delta \Omega}{2}), \\
F_{\Delta} & = &
m_\Delta \bar{\Delta} + 
a (\Omega \bar{\Delta} - \frac{\Tr \Omega \bar{\Delta}}{2}), \\
F_{\Omega} & = &
m_\Omega \Omega + 
a (\bar{\Delta} \Delta - \frac{\Tr \bar{\Delta} \Delta}{2}) +
\alpha (\Phi_1 \tau_2 \Phi_2^T \tau_2 - \Phi_2 \tau_2 \Phi_1^T \tau_2), \\
F_{\bar{\Delta_c}} & = &
m_\Delta \Delta_c + 
a (\Delta_c \Omega_c - \frac{\Tr \Delta_c \Omega_c}{2}), \\
F_{\Delta_c} & = &
m_\Delta \bar{\Delta_c} + 
a (\Omega_c \bar{\Delta_c} - \frac{\Tr \Omega_c \bar{\Delta_c}}{2}), \\
F_{\Omega_c} & = &
m_\Omega \Omega_c + 
a (\bar{\Delta_c} \Delta_c - \frac{\Tr \bar{\Delta_c} \Delta_c}{2}) +
\alpha (\Phi_1^T \tau_2 \Phi_2 \tau_2 - \Phi_2^T \tau_2 \Phi_1 \tau_2), \\
F_{\Phi_1} & = &
2 \mu_{11} \tau_2 \Phi_1^T \tau_2 +
2 \mu_{12} \tau_2 \Phi_2^T \tau_2 +
\alpha (
\tau_2 \Phi_2^T \tau_2 \Omega - 
\tau_2 \Phi_2^T \Omega^T \tau_2 +
\tau_2 \Omega_c \Phi_2^T \tau_2 -
\Omega_c^T \tau_2 \Phi_2^T \tau_2
), \\
F_{\Phi_2} & = &
2 \mu_{12} \tau_2 \Phi_1^T \tau_2 +
2 \mu_{22} \tau_2 \Phi_2^T \tau_2 -
\alpha (
\tau_2 \Phi_1^T \tau_2 \Omega - 
\tau_2 \Phi_1^T \Omega^T \tau_2 +
\tau_2 \Omega_c \Phi_1^T \tau_2 -
\Omega_c^T \tau_2 \Phi_1^T \tau_2
).
\end{array}
\end{equation}
After substituting the vevs, the expressions for the F-terms become
\begin{equation}
\label{eq:FTermVevs}
\begin{array}{rcl}
\vev{F_{\bar{\Delta}}} & = &
\left(
\begin{array}{c c}
0 & 0 \\
d(m_\Delta + a \omega) & 0
\end{array}
\right), \\
\vev{F_{\Delta}} & = &
\left(
\begin{array}{c c}
0 & \bar{d}(m_\Delta + a \omega) \\
0 & 0
\end{array}
\right), \\
\vev{F_{\Omega}} & = &
\left(
\begin{array}{c c}
m_\Omega \omega + \frac{a d \bar{d}}{2} 
+ \alpha(k_1 k_2' - k_1' k_2) & 
0 \\
0 & 
-(m_\Omega \omega + \frac{a d \bar{d}}{2} 
  + \alpha(k_1 k_2' - k_1' k_2))
\end{array}
\right), \\
\vev{F_{\bar{\Delta}_c}} & = &
\left(
\begin{array}{c c}
0 & d_c(m_\Delta + a \omega_c) \\
0 & 0
\end{array}
\right), \\
\vev{F_{\Delta_c}} & = &
\left(
\begin{array}{c c}
0 & 0 \\
\bar{d}_c(m_\Delta + a \omega_c) & 0
\end{array}
\right), \\
\vev{F_{\Omega_c}} & = &
\left(
\begin{array}{c c}
m_\Omega \omega_c 
+ \frac{a d_c \bar{d}_c}{2} + \alpha(k_1 k_2' - k_1' k_2) & 
0 \\
0 & 
-(m_\Omega \omega_c 
+ \frac{a d_c \bar{d}_c}{2} + \alpha(k_1 k_2' - k_1' k_2))
\end{array}
\right), \\
\vev{F_{\Phi_1}} & = &
\left(
\begin{array}{c c}
2 \mu_{11} k_1' + 2 \mu_{12} k_2' + 2 \alpha k_2' (\omega - \omega_c) &0 \\
0& 2 \mu_{11} k_1 + 2 \mu_{12} k_2 - 2 \alpha k_2 (\omega - \omega_c)    \\
\end{array}
\right), \\
\vev{F_{\Phi_2}} & = &
\left(
\begin{array}{c c}
2 \mu_{12} k_1' + 2 \mu_{22} k_2' - 2 \alpha k_1' (\omega - \omega_c) &0 \\
0& 2 \mu_{12} k_1 + 2 \mu_{22} k_2 + 2 \alpha k_1 (\omega - \omega_c)    \\
\end{array}
\right).
\end{array}
\end{equation}

We now investigate what vevs ensure flatness conditions for all the F-terms and all the D-terms.
For generic values of $\mu_{11}$, $\mu_{12}$ and $\mu_{22}$, the
entries of $F_{\Phi_1}$ and $F_{\Phi_2}$ are linearly independent.
Hence flatness of $F_{\Phi_1}$ and $F_{\Phi_2}$ implies that
the vevs of $k_1$, $k'_1$, $k_2$, $k'_2$ are all zero.
This automatically makes $D_{L,3}$ and $D_{R,3}$ flat. 

Proceeding ahead, we now see that the F-flatness conditions split into
two subsets viz. the left handed conditions and right handed conditions.
This allows to conclude, as noted first by Aulakh et al., that the
complete solution set for F-flatness and D-flatness is obtained by taking 
$(\omega, d) = (0, 0)$ or 
$(\omega, d) = 
(\frac{m_\Delta}{-a}, \frac{\sqrt{2 m_\Omega m_\Delta}}{-a})$,
and
$(\omega_c, d_c) = (0, 0)$ or 
$(\omega_c, d_c) = 
(\frac{m_\Delta}{-a},\frac{\sqrt{2 m_\Omega m_\Delta}}{-a})$
and
$(k_1, k'_1, k_2, k'_2) = (0, 0, 0, 0)$ (note $a < 0$).
Thus, the complete SUSY preserving solution set has,
beside the trivial all zero solution, three non-trivial solutions.
Of these, the solution
$
(\omega, d, \omega_c, d_c, k_1, k_1', k_2, k_2') =
(\frac{m_\Delta}{-a}, \frac{\sqrt{2 m_\Omega m_\Delta}}{-a}, 
 \frac{m_\Delta}{-a}, \frac{\sqrt{2 m_\Omega m_\Delta}}{-a}, 0,0,0,0)
$
is unphysical because it breaks down the gauge symmetry to 
$\SU(3)_c$ even though it preserves SUSY. The other two solutions are
indeed physical and can be interpreted as breaking down the left-right
symmetric gauge group 
$\SU(3)_c \times \SU(2)_L \times \SU(2)_R \times U(1)_{B-L}$ 
into either the left-handed Minimal Supersymmetric Standard Model (MSSM)
$\SU(3)_c \times \SU(2)_L \times U(1)_Y$, 
or into the right-handed MSSM 
$\SU(3)_c \times \SU(2)_R \times U(1)_Y$. 

The expressions for the vevs must be related to the two physical scales in LRSUSY. 
We need to set 
\begin{eqnarray}
M_R &\cong& \frac{m_\Delta}{-a} \\
M_{B-L} &\cong& \frac{\sqrt{2 m_\Omega m_\Delta}}{-a}
\end{eqnarray}
in the LH regions and likewise in the RH regions. A solution to the proliferation of mass scales was
sought in \cite{Aulakh:1997fq} by invoking an $R$ symmetry of the superpotential which 
forbids the terms $\Omega^2$ and $\Omega_c^2$. The $R$ charge values can be set to
\begin{eqnarray}
\Delta, \bar{\Delta}, \Delta_c,  \bar{\Delta_c},  \Phi_i & \rightarrow  & 1  \\
\Omega  & \rightarrow  &  0 \\
L, L_c, Q, Q_c & \rightarrow  & \frac{1}{2}  
\end{eqnarray}
where the $L$, $Q$ etc are matter superfields which are not relevant to this paper.
The terms $\Omega^2$ and $\Omega_c^2$ can then be introduced only as
soft terms, with the coefficients $m_\Omega=m_{\Omega_c}$ determined by SUSY breaking scale $\cong M_{EW}$.
This leads to an elegant simplification giving rise to the see-saw relation  
\begin{equation}
\label{eq:BLscaleseesaw}
M_{B-L}^2 \cong M_R M_{EW}
\end{equation}

From the cosmological viewpoint,  the early universe enters an epoch with two types of domains.
In the left handed (LH) domains, the right handed vevs take non-zero values
and in the right handed (RH) domains, the left handed vevs take non-zero values.
The corresponding SUSY preserving vevs are
\begin{equation}
\label{eq:domainvevs}
\begin{array}{r c l}
(\omega, d, \omega_c, d_c, k_1, k_1', k_2, k_2') 
& = &
(0, 0, \frac{m_\Delta}{-a}, \frac{\sqrt{2 m_\Omega m_\Delta}}{-a}, 0,0,0,0)
~~~
\mbox{LH domain}, \\
(\omega, d, \omega_c, d_c, k_1, k_1', k_2, k_2') 
& = &
(\frac{m_\Delta}{-a}, \frac{\sqrt{2 m_\Omega m_\Delta}}{-a}, 0, 0, 0,0,0,0)
~~~
\mbox{RH domain}.
\end{array}
\end{equation}

The formation of the two types of domains also leads to topological domain walls 
separating them. This is because together with the breaking of the gauge
symmetry, a discrete left-right symmetry is also broken. The presence
of these walls or energy barriers conflicts with current cosmology. Several
earlier works have discussed how such walls may be made to disappear
fast enough so as to be consistent with present day observations
\cite{Mishra}. 
The older study \cite{Anjishnu} demonstrated the existence of domain walls
containing a CP violating phase  in LRSUSY with implications to leptogenesis, 
but only as a proof-of-concept study. 
In this paper, we extend their ideas greatly and
come up with quantitative estimates relating the parameter ranges that can give rise to
the required spatially varying CP  violating phase within the domain wall with the 
non-zero phase required for electron EDM in zero temperature translation invariant theory.
The next section provides the details.

To the SUSY scalar potential 
\begin{equation}
\label{eq:VSUSY}
\begin{array}{r c l}
V_{\mathrm{SUSY}} 
& = &
|F_{\bar{\Delta}}|^2 +
|F_{\Delta}|^2 +
|F_{\bar{\Delta}_c}|^2 +
|F_{\Delta_c}|^2 +
|F_{\Omega}|^2 +
|F_{\Omega_c}|^2 +
|F_{\Phi_1}|^2 +
|F_{\Phi_2}|^2 \\
&  &
{} +
\sum_{m=1}^3 (|D_{L,m}|^2 + |D_{R,m}|^2) +
|D_{B-L}|^2,
\end{array}
\end{equation}
we add the following soft mass terms for the bidoublets
\begin{equation}
\label{eq:Vsoft}
\begin{array}{r c l}
V_{\mathrm{soft}} 
& = &
-\mu_1^2 \Tr (\Phi_1^\dag \Phi_1) 
-\mu_2^2 \Tr (\Phi_2^\dag \Phi_2) \\
&   &
{}
-e^{i\beta_3} \mu_3^2 \Tr (\Phi_1^\dag \tau_2 \Phi_1^* \tau_2) 
-e^{i\beta_4} \mu_4^2 \Tr (\Phi_2^\dag \tau_2 \Phi_2^* \tau_2) 
-e^{i\beta_5} \mu_5^2 \Tr (\Phi_1^\dag \tau_2 \Phi_2^* \tau_2) 
+ \mathrm{h.c.},
\end{array}
\end{equation}
where $\mu_i^2 > 0$ and $\beta_3$, $\beta_4$, $\beta_5$ 
are explicit CP phases.
Substituting the vevs we get
$
\vev{V} =
\vev{V_{\mathrm{SUSY}}} +
\vev{V_{\mathrm{soft}}},
$
where
\begin{eqnarray*}
\vev{V_{\mathrm{SUSY}}} 
& = &
|\vev{F_{\bar{\Delta}}}|^2 +
|\vev{F_{\Delta}}|^2 +
|\vev{F_{\bar{\Delta}_c}}|^2 +
|\vev{F_{\Delta_c}}|^2 +
|\vev{F_{\Omega}}|^2 +
|\vev{F_{\Omega_c}}|^2 +
|\vev{F_{\Phi_1}}|^2 +
|\vev{F_{\Phi_2}}|^2 \\
&   &
{} +
|\vev{D_{L,3}}|^2 + |\vev{D_{R,3}}|^2 + |\vev{D_{B-L}}|^2,
\end{eqnarray*}
and
\begin{equation}
\label{eq:VSoftVevs}
\begin{array}{rcl}
\vev{V_{\mathrm{soft}}} 
& = &
-\mu_1^2 (|k_1|^2 + |k'_1|^2) 
-\mu_2^2 (|k_2|^2 + |k'_2|^2) \\
&  &
{}
- 4 \mu_3^2 \mathrm{Re}(e^{i\beta_3} k_1^* (k'_1)^*)
- 4 \mu_4^2 \mathrm{Re}(e^{i\beta_4} k_2^* (k'_2)^*)
- 2 \mu_5^2 \mathrm{Re}(e^{i\beta_5} (k_1^* (k'_2)^* + (k'_1)^* k_2^*)).
\end{array}
\end{equation}
We shall take the fine tuning condition 
$\mu_{12}^2 \approx \mu_{11} \mu_{22} + \alpha^2 M_R^2$ of Aulakh et
el. \cite{Aulakh:1997fq} which ensures that out of the four neutral Higgs scalars
that arise from the bidoublets after the breaking of 
$\SU(2)_R \times \U(1)_{B-L}$ symmetry, two of them have masses near
zero (the other two end up having mass near $M_R$). 
Our soft masses $\mu_1, \ldots, \mu_5$ 
are chosen to be around $\alpha^2 M_R$, so that at temperatures
below the SUSY breaking scale, the two light
Higgs scalars eventually become the Higgs bosons of a two Higgs doublet
model (2HDM) satisfying $\SU(3)_c \times \SU(2)_L \times U(1)_Y$ gauge
symmetry. At even lower temperatures, the SM Higgs boson arises from
the 2HDM.

\section{Spatially varying Higgs vevs}
The SUSY scalar potential and the soft mass terms for the
bidoublets receive temperature corrections determined by the scale $M_{B-L}$.
The temperature dependence of the squared mass term of a Higgs scalar has
been evaluated at the one loop level in earlier works
\cite{Dolan}.
For this indicative study we shall take the temperature correction to each
mass matrix element to be\cite{Cline2HDM,AndersonHall,RamseyEWBG},
\begin{equation}
(\Delta m^2)^T \sim O(g^2 T^2)
\end{equation}  
The full temperature dependent mass matrix can be found in
Appendix~\ref{sec:tempmass}. Thus,
going from temperature $M_{B-L}$ to zero temperature entails the
lowering of the mass matrix elements by $O(g^2 T^2)$. 
For the leptogenesis calculations, we work at the high
temperature  $T=M_{B-L}$ with the mass matrix arising from the SUSY scalar
potential and the soft masses described above.
These choices for the mass parameters ensure that the mass matrix
of the bidoublets has a negative eigenvalue inside the wall whereas
all its eigenvalues are positive outside. This in turn means that it is energetically 
favourable for the bidoublet fields to take non-zero vevs inside the wall while 
continuing to take zero vevs outside, even though the soft terms have negative squared
masses. For the zero temperature electron EDM calculation that we do later, 
we can work in a four Higgs doublet model
\cite{Cline2HDM,Fromme} with temperature corrections dropped from the 
mass matrix of Appendix~\ref{sec:tempmass}.

In this section we shall see that,
for a certain choice of soft SUSY breaking mass terms for the bidoublets,
it becomes energetically favourable for the bidoublet fields to take
non-zero vevs within the wall while continuing to take (almost) zero vevs 
outside it. 
Moreover, the generic $O(1)$ phases in the soft mass terms entail a consequence that 
the bidoublet fields take on spatially  varying $CP$ violating phases inside the wall 
while returning to constant non-zero phases outside the wall.

Let the spatial coordinate giving the distance from
the wall, in units of inverse temperature $1/T$, be denoted by $x$. 
The fields constituting the wall have substantial variation 
only in a narrow region $\Delta x \sim (L /v)$ where $v$ is a generic 
scalar vacuum expectation value and $L^{-1}\sim \sqrt{\lambda}$ is derived from 
a generic quartic coupling $\lambda$ of a renormalisable field theory.
The SUSY preserving vev 8-tuples inside the two domains are given by 
Equation~\ref{eq:domainvevs}.
We want a smooth 
variation of the vev 8-tuple as a function of $x$, going from the 
LH domain to the RH domain passing
through the wall on the way. By the argument in the previous section,
this necessarily entails breaking SUSY inside the wall. Thus in order
to get the shapes of the $x$-dependent vevs of the Higgs fields, we need
to first write down a functional for the energy per unit area of the
wall and then minimise the functional via Euler-Lagrange equations.

Let $\dot{f}$ denote the derivative of vev of field $f$ with respect to
$x$. Let $r_1$, $i_1$ be the real and imaginary parts of vev of $k_1$,
$r_2, i_2, \ldots, r'_2, i'_2$ the real and imaginary parts of the vevs of the
respective bidoublet fields. We make the simplifying assumptions that
the non-bidoublet Higgs fields are real everywhere. 
The finite temperature energy per unit area, 
which is the sum of gradient energies and potential energies of all the fields, 
plus field dependent temperature corrections can now be taken to be, 
\begin{equation}
\label{eq:EDensity}
\begin{array}{rcl}
H^T
& = &
\int \, dx \,
(
\frac{1}{2} (
\dot{\omega}^2 +
\dot{\omega_c}^2 +
\dot{d}^2 +
\dot{\bar{d}}^2 +
\dot{d_c}^2 +
\dot{\bar{d}_c}^2  \\
&  &
~~~~~~~~~~~
+ \dot{r_1}^2 +
\dot{i_1}^2 +
\dot{r'_1}^2 +
\dot{i'_1}^2 +
\dot{r_2}^2 +
\dot{i_2}^2 +
\dot{r'_2}^2 +
\dot{i'_2}^2 
) \\
&  &
~~~~~~~~~~~
+ \vev{V_{\mathrm{SUSY}}} + \vev{V^T_{\mathrm{soft}}}).
\end{array}
\end{equation}
Here a superscript $T$ on $H$ and $V_{\mathrm{soft}}$ is a reminder of the temperature dependence.
We determine the domain wall solutions such that SUSY is preserved asymptotically by the vevs 
upto relatively small temperature correction. It is only in the narrow region of the wall where the 
omega fields become small that the temperature dependent terms and soft terms become more significant. 
In the equations below, all the vev are meant to be temperature dependent, though for 
simplicity of notation we drop the superscript $T$. 
\begin{equation}
\label{eq:LeftVevs}
\begin{array}{c}
\vev{\omega(-\infty)} = 
\vev{d(-\infty)} = 
\vev{\bar{d}(-\infty)} = 0, \\
\vev{\omega_c(-\infty)} = M_R, 
\vev{d_c(-\infty)} = 
\vev{\bar{d}_c(-\infty)} = M_{B-L}, \\
\vev{r_1(-\infty)} = 
\vev{i_1(-\infty)} = 
\vev{r'_1(-\infty)} = 
\vev{i'_1(-\infty)} = 0, \\ 
\vev{r_2(-\infty)} = 
\vev{i_2(-\infty)} = 
\vev{r'_2(-\infty)} = 
\vev{i'_2(-\infty)} = 0
\end{array}
\end{equation}
in the left domain and
\begin{equation}
\label{eq:RightVevs}
\begin{array}{c}
\vev{\omega_c(\infty)} = 
\vev{d_c(\infty)} = 
\vev{\bar{d}_c(\infty)} = 0, \\
\vev{\omega(-\infty)} = M_R, 
\vev{d(\infty)} = 
\vev{\bar{d}(\infty)} = M_{B-L}, \\
\vev{r_1(\infty)} = 
\vev{i_1(\infty)} = 
\vev{r'_1(\infty)} = 
\vev{i'_1(\infty)} = 0, \\ 
\vev{r_2(\infty)} = 
\vev{i_2(\infty)} = 
\vev{r'_2(\infty)} = 
\vev{i'_2(\infty)} = 0
\end{array}
\end{equation}
in the right domain.


With all the parameters in place, we can now
minimise the energy density per unit area by solving the Euler-Lagrange
equations arising from Equation~\ref{eq:EDensity}. 
The equations are
given explicitly in Appendix~\ref{sec:EL} for completeness.
In the next section
we describe how to solve them numerically in order to obtain the
shapes of the spatially varying vevs of the Higgs fields both within and
outside the wall. The vevs take the limiting values described in
Equations~\ref{eq:LeftVevs}, \ref{eq:RightVevs} outside the wall.

\section{Solving the Euler-Lagrange equations for Higgs vevs}
The Euler-Lagrange equations in Appendix~\ref{sec:EL} 
form a coupled
system of second order non-linear differential equations satisfying
the boundary conditions of Equation~\ref{eq:LeftVevs} 
in the left domain and Equation~\ref{eq:RightVevs} 
in the right domain. Since the
derivatives of the vevs are zero in both left and right domains, these
equations are not well-suited for numerical solution by shooting
methods. Because of the large number of non-linear equations, 
their numerical solution also
faces difficulties under finite element or path deformation
\cite{Wainwright} methods.

Naive attempts to solve the Euler-Lagrange equations as an initial
value system also run into problems. This is 
because if we take the initial conditions at a
point in the left domain, the algorithms give us the SUSY flat 
left domain solution only as that solution minimises the energy density
to zero. A similar statement holds if we take the initial conditions at a
point in the right domain. We have to somehow model the loss of 
translation invariance due to the domain wall in our solution.

Since $\Omega$, $\Omega_c$ have the heaviest vevs 
outside the wall, we 
fix a natural ansatz for them that smoothly goes from the LH solution
to the RH solution while passing through the wall on the way. The
wall is assumed to extend from $-L$ to $L$ in units of inverse
temperature. The 
ansatz takes the form of kink functions:
\begin{equation}
\label{eq:OmegaAnsatz}
\begin{array}{rcl}
\omega_c(x) 
& = &
(1 - \tanh(\frac{m_\Delta }{-2aL}x)) \frac{m_\Delta}{-2a}, \\
\omega(x) 
& = &
(1 + \tanh(\frac{m_\Delta }{-2aL}x)) \frac{m_\Delta}{-2a}. \\
\end{array}
\end{equation}
The ansatz has the property that $\omega(x)$, $\omega_c(x)$ take the
correct limiting values outside the wall in both domains,
but are non-zero within the wall. In the example plots later on,
we shall be taking $L\sim (\sqrt{\lambda})^{-1}\sim 5$ for 
concreteness.

Fixing the ansatz for the vevs of $\Omega$, $\Omega_c$ models the
effect of the wall and reduces the
Euler-Lagrange equations to a set of 12 coupled second order
differential equations for the four triplet Higgs vevs $d$, $\bar{d}$,
$d_c$, $\bar{d}_c$ and the eight vevs corresponding to the real and 
imaginary parts of the bidoublet Higgs fields. 

With this setting we solve the Euler-Lagrange equations as an initial
value problem
numerically using the GSL 2.6 library, setting the stepping function
to be Runge-Kutta Dormand-Prince (8,9) with step size, absolute error
and relative error of $10^{-6}$. 
The obtained solutions for the bidoublet vevs are non-zero 
and spatially varying
inside the wall but become zero outside. For the triplet vevs the
obtained solutions are spatially varying inside the wall and approach
their constant SUSY determined values outside. 
The vevs of $\Delta$, $\Delta_c$ are very sensitive to the values of
the vevs of $\Omega$, $\Omega_c$ and quickly drop to 
zero towards the centre of the wall as that minimises
the energy density.

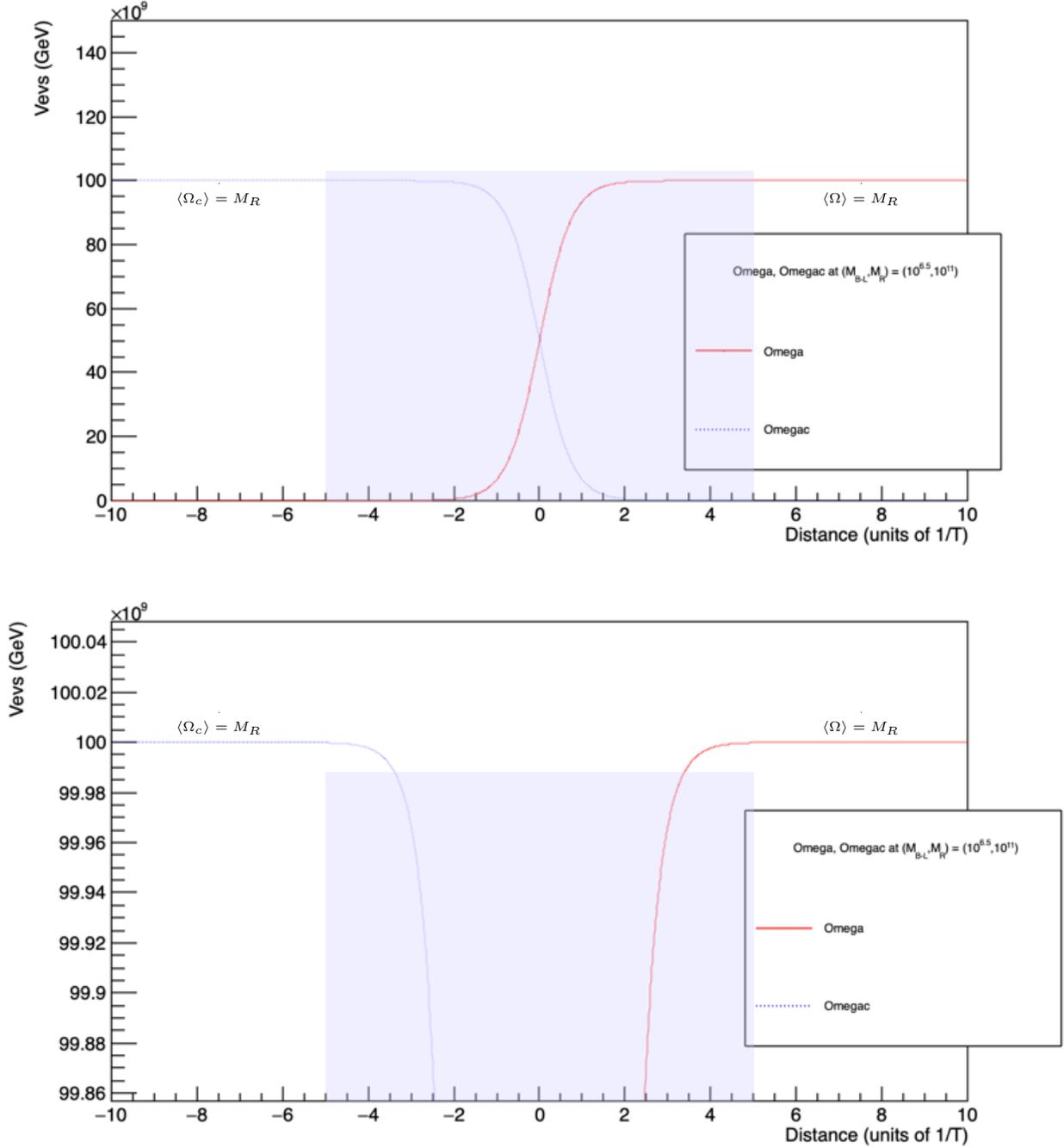
\begin{figure}[!hhh]
\begin{tikzpicture}
\node[anchor=south west,inner sep=0] (image) at (0,0) 
{\includegraphics[width=\textwidth]{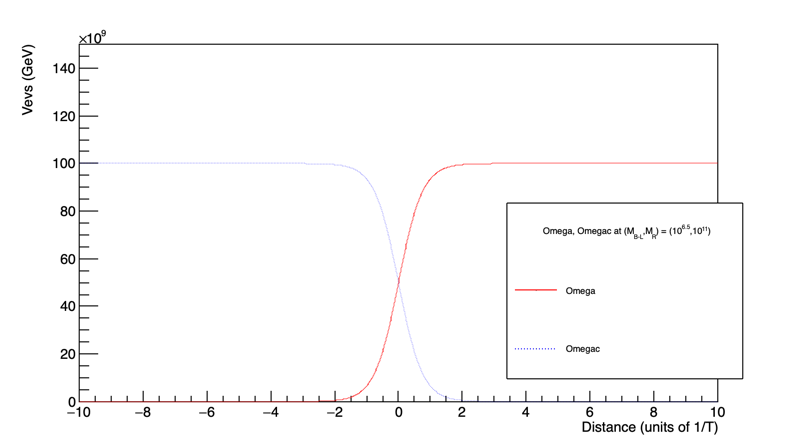}};
\begin{scope}[x={(image.south east)},y={(image.north west)}]
\fill[LBlue,opacity=0.3] (0.3,0.1) rectangle (0.7,0.65);
\draw (0.2, 0.63) -- (0.2, 0.63) node[anchor=north] 
{{\tiny $\vev{\Omega_c} = M_R$}};
\draw (0.8, 0.63) -- (0.8, 0.63) node[anchor=north] 
{{\tiny $\vev{\Omega} = M_R$}};
\end{scope}
\end{tikzpicture}

\begin{tikzpicture}
\node[anchor=south west,inner sep=0] (image) at (0,0) 
{\includegraphics[width=\textwidth]{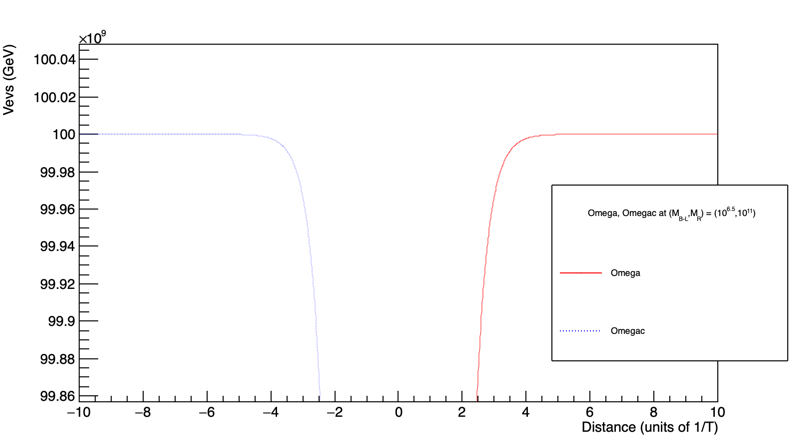}};
\begin{scope}[x={(image.south east)},y={(image.north west)}]
\fill[LBlue,opacity=0.3] (0.3,0.1) rectangle (0.7,0.65);
\draw (0.2, 0.75) -- (0.2, 0.75) node[anchor=north] 
{{\tiny $\vev{\Omega_c} = M_R$}};
\draw (0.8, 0.75) -- (0.8, 0.75) node[anchor=north] 
{{\tiny $\vev{\Omega} = M_R$}};
\end{scope}
\end{tikzpicture}
\caption{
{\bf Above:}
Plots of  vacuum expectation values $\omega$, $\omega_c$
as a function of the distance $x$, in units of $1/T$, from the 
domain wall, plotted for 
$T = M_{B-L} = 10^{6.5}~\mathrm{GeV}$,
$M_R = 10^{11}~\mathrm{GeV}$, $L = 5$, 
$\lambda \sim L^{-2} = 0.04$ and
the choice of the other parameters,
$a = -1.5$, $\alpha = 0.006$,
$\mu_{11} = 0.7 M_R$, $\mu_{22} = 0.7 M_R$, 
$\mu_1 = 0.2 \alpha^2 M_R$,
$\mu_2 = 0.3 \alpha^2 M_R$,
$\mu_3 = 0.5 \alpha^2 M_R$, $\beta_3 = -0.5$,
$\mu_4 = 1.5 \alpha^2 M_R$, $\beta_4 = 1.2$,
$\mu_5 = 0.1 \alpha^2 M_R$, $\beta_5 = -1.4$. 
The domain wall stretches from $-L$ to $L$.
{\bf Below:} The same plot magnfied, showing the gradual drop of the
vevs of $\omega$, $\omega_c$ just inside the wall around $x = \pm L$.
}
\label{fig:Triplets-Omega}
\end{figure}

\begin{figure}[!hhh]
\begin{tikzpicture}
\node[anchor=south west,inner sep=0] (image) at (0,0) 
{\includegraphics[width=\textwidth]{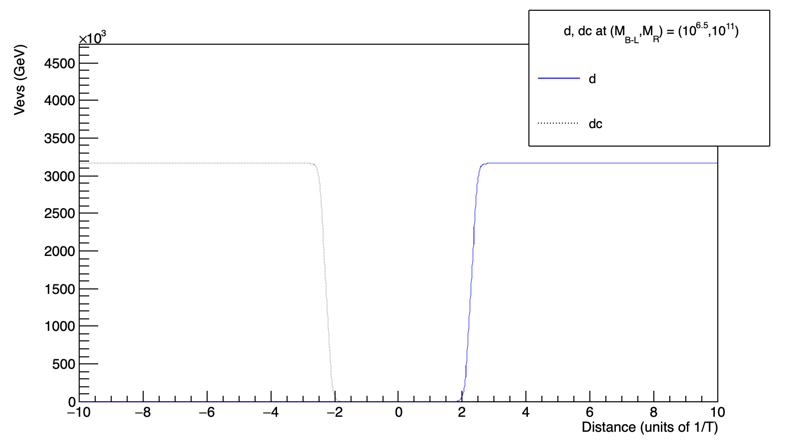}};
\begin{scope}[x={(image.south east)},y={(image.north west)}]
\fill[LBlue,opacity=0.3] (0.3,0.1) rectangle (0.7,0.65);
\draw (0.2, 0.63) -- (0.2, 0.63) node[anchor=north] 
{{\tiny $\vev{\Delta_c} = M_{B-L}$}};
\draw (0.8, 0.63) -- (0.8, 0.63) node[anchor=north] 
{{\tiny $\vev{\Delta} = M_{B-L}$}};
\end{scope}
\end{tikzpicture}
\caption{Plots of  vacuum expectation values $d$, $d_c$ obtained for 
$T = M_{B-L} = 10^{6.5}~\mathrm{GeV}$,
$M_R = 10^{11}~\mathrm{GeV}$, and for the ansatz and 
and parameters as in Fig. \ref{fig:Triplets-Omega}.}
\label{fig:Triplets-Delta}
\end{figure}

Figures~\ref{fig:Triplets-Omega} and \ref{fig:Triplets-Delta} show how the 
vacuum expectation values of the heavy triplet Higgs fields 
$\Omega$, $\Omega_c$, $\Delta$, $\Delta_c$ vary as a function of the 
distance $x$ from
the wall for an ad hoc setting of parameters 
$a = -1.5$, $\alpha = 0.006$, 
$\mu_{11} = 0.7 M_R$, $\mu_{22} = 0.7 M_R$, 
$\mu_1 = 0.2 \alpha^2 M_R$,
$\mu_2 = 0.3 \alpha^2 M_R$,
$\mu_3 = 0.5 \alpha^2 M_R$, $\beta_3 = -0.5$,
$\mu_4 = 1.5 \alpha^2 M_R$, $\beta_4 = 1.2$,
$\mu_5 = 0.1 \alpha^2 M_R$, $\beta_5 = -1.4$,
$M_R = 10^{11}~\mathrm{GeV}$ and $M_{B-L} = 10^{6.5}~\mathrm{GeV}$.
Finally the generic quartic coupling $\lambda$ which can be determined 
from 
those appearing in the potential, is taken to be 
$\lambda \sim L^{-2} = 0.04$. 

\begin{figure}[!hhh]
\begin{tikzpicture}
\node[anchor=south west,inner sep=0] (image) at (0,0) 
{\includegraphics[width=\textwidth]{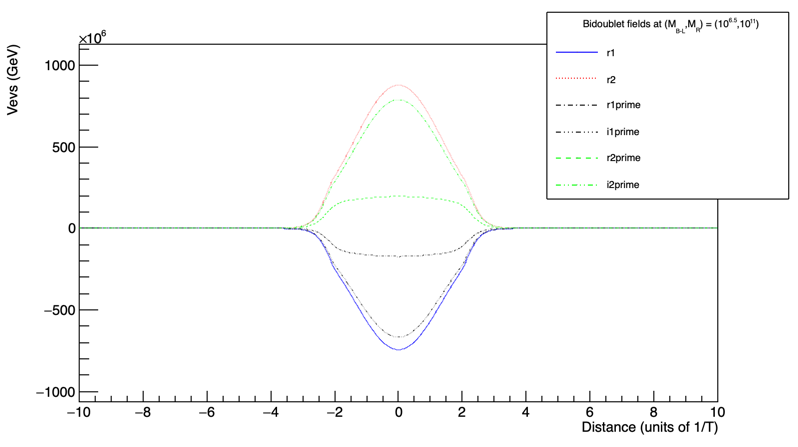}};
\begin{scope}[x={(image.south east)},y={(image.north west)}]
\fill[LBlue,opacity=0.3] (0.3,0.1) rectangle (0.7,0.85);
\draw (0.45, 0.8) -- (0.45, 0.8) node[anchor=north east] 
{{\tiny $\vev{\Phi}$}};
\end{scope}
\end{tikzpicture}
\caption{{\bf Left:} Real and imaginary parts of the bidoublet Higgs
fields, $r_1$, $r'_1$, $i'_1$, $r_2$, $r'_2$, $i'_2$,
as a function of the distance $x$, in units of $1/T$, from the 
domain wall, plotted for $T = M_{B-L} = 10^{6.5}~\mathrm{GeV}$,
$M_R = 10^{11}~\mathrm{GeV}$, and other parameters as in 
Fig. \ref{fig:Triplets-Omega}.
}
\label{fig:Phi-vevs}
\end{figure}

\begin{figure}[!hhh]
\begin{tikzpicture}
\node[anchor=south west,inner sep=0] (image) at (0,0) 
{\includegraphics[width=\textwidth]{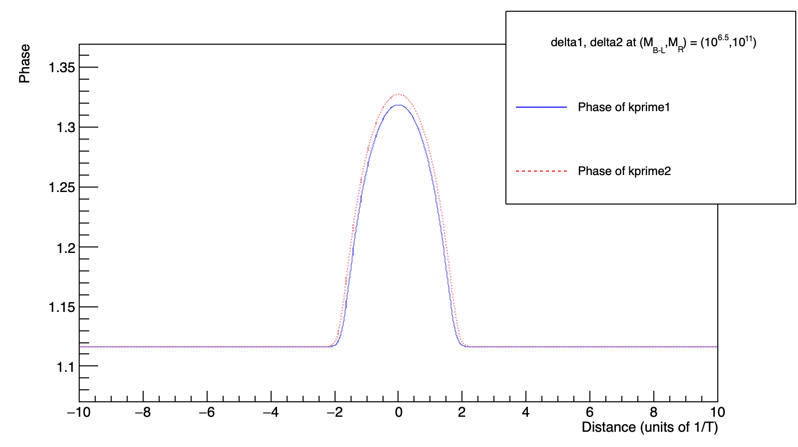}};
\begin{scope}[x={(image.south east)},y={(image.north west)}]
\fill[LBlue,opacity=0.3] (0.3,0.1) rectangle (0.7,0.88);
\draw (0.48, 0.8) -- (0.48, 0.8) node[anchor=north east] 
{{\tiny $\delta_{\mathrm{CP}}$}};
\end{scope}
\end{tikzpicture}
\caption{The phases of $k'_1$, $k'_2$ plotted as a function of $x$
for the same parameters as in Fig. \ref{fig:Triplets-Omega} and 
Fig. \ref{fig:Phi-vevs}.}
\label{fig:Phi-phases}
\end{figure}

Figure~\ref{fig:Phi-vevs} exhibits the spatial variation of the real and
imaginary parts of the vacuum expectation values of the bidoublets. It 
turns out that the vevs of the fields
$k_1$ and $k_2$ are real throughout while $k'_1$ and $k'_2$ do take
spatially varying complex vevs.  In Figure~\ref{fig:Phi-phases}, we plot 
the complex phases of vevs of $k'_1$ and $k'_2$ as a function of $x$. 
Observe
that the phases vary inside the wall but converge to a constant non-zero
value outside.

These considerations show that a spatially varying CP violating phase 
can indeed be produced by the bidoublet Higgs vevs inside the domain 
wall in the early universe.
This phase persists as a constant non-zero quantity outside the wall.
In the next two sections, we investigate the implications of this
phenomenon for the 
electric dipole moment of the electron and the
baryon asymmetry of the universe.

\section{Electron EDM constraints on LRSUSY}
The zero temperature mass matrix of the neutral components
of the bidoublet Higgs fields is given in Appendix~\ref{sec:tempmass}.  
As there are two neutral complex 
components in each bidoublet, we get in total eight real neutral
fields and so the mass matrix is $8 \times 8$. 
It turns out that the neutral mass eigenstates induce complex phases
for the bidoublet Higgs fields
relative to each other. This is true both within the wall as well
as outside it. This feature gives rise to an
electric dipole moment (EDM) for the electron
at one loop and two loop levels. The maximum effect on the electron
EDM is exerted by the lightest mass eigenstate. 

\begin{figure}[!hhh]
{\color{yellow} \rule{\textwidth}{2pt}}

\begin{tikzpicture} 
\begin{feynman}
\vertex (e1); 
\vertex [right=of e1] (e2); 
\node [above right=of e2, crossed dot] (h1);
\vertex [above right=of h1] (l1);
\vertex [above right=of l1] (l2);
\vertex [above=of l2] (p1);
\vertex [below right=of l2] (l3);
\vertex [below right=of l3] (h2);
\vertex [right=of e2] (e3);
\vertex [right=of e3] (e4); 
\vertex [right=of e4] (e5); 
\vertex [right=of e5] (e6); 
\vertex [right=of e6] (e7); 
\diagram* {
  (e1) -- [fermion, edge label'=\(e^-\)] 
     (e2) --  (e3)  --  (e4) -- [fermion, edge label'=\(e^-\)] 
     (e5) -- 
     (e6) -- [fermion, edge label'=\(e^-\)] (e7),
  (e2) -- [scalar, edge label=\(H_0\)] 
    (h1) -- [scalar] (l1), 
    (l1) -- (l2) -- (l3) -- [edge label=\(\mbox{$W^-, t$}\)] (l1),
    (l3) -- [boson, edge label=\(\mbox{$\gamma, Z$}\)] (e6),
  (p1) -- [boson, edge label'=\(\gamma\)] (l2),
}; 
\end{feynman}
\end{tikzpicture}
\caption{
The two loop diagram giving the maximum contribution to the electron EDM
in our model. Four such diagrams have to be calculated, corresponding
to the choice of top quark or $W$ boson
in the inner loop, and the choice of $H_i \gamma \gamma$ or $H_i Z \gamma$
as the bosons interacting with this inner  loop. Here $H_0$ denotes
the lightest mass eigenstate of the bidoublet Higgs.
}
\label{fig:twoloopedm}

{\color{yellow} \rule{\textwidth}{2pt}}
\end{figure}
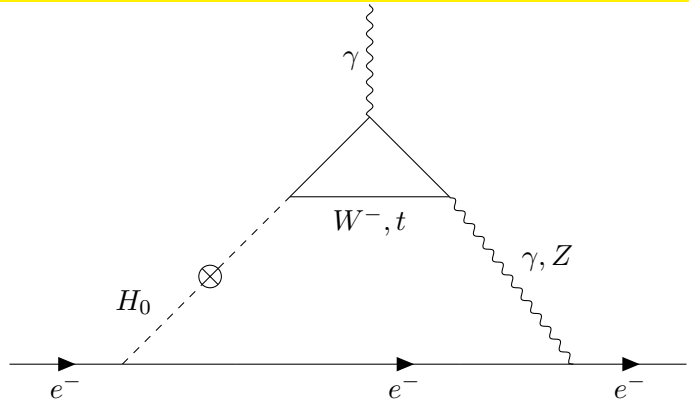

The one loop contribution to electron EDM $d_e$ 
is given by \cite{RamseyEWBG}
\begin{equation}
\label{eq:EDMoneloop}
(d_e/e) |_{\mathrm{one~loop}} \sim
\frac{\alpha m_e}{4 \pi M_h^2} \sin \delta,
\end{equation}
where $M_h$ is the mass of the lightest eigenstate of the bidoublet
Higgs, $\alpha$ is the fine structure constant evaluated 
at the scale $M_h$ and $\delta$ is the complex relative phase between
the neutral scalars induced by the lightest mass eigenstate.
Surprisingly however, for large values of $M_{B-L}$ and $M_R$ two loop
effects arising from the neutral scalars dominate the one loop effect.
This was first realised by Barr and Zee \cite{BarrZee} and then refined
by several other authors. 
We use the formulas of Chang, Keung and Yuan \cite{Chang}
in order to compute the two loop contribution. The two loop contribution
is a sum of contributions from several diagrams. Most important amongst
those are four diagrams coming from Figure~\ref{fig:twoloopedm} 
arising from 
the choice of top quark or $W$ boson
in the inner loop, and the choice of $H \gamma \gamma$ or $H Z \gamma$
as the bosons interacting with this inner  loop. 
Their total contribution is of the form
\begin{equation}
\label{eq:EDMtwoloop}
\begin{array}{r c l}
(d_e/e) |_{\mathrm{two~loop}} 
& = &
\frac{G_F m_e \alpha \sin \delta}{\pi^3 \sqrt{2}}
(
f_{W, H \gamma \gamma}(M_W^2/M_h^2) +
f_{W, H Z \gamma}(M_W^2/M_h^2) \\
&  &
~~~~~~~~~~~~~~
{} +
f_{t, H \gamma \gamma}(M_t^2/M_h^2) +
f_{t, H Z \gamma}(M_t^2/M_h^2) 
).
\end{array}
\end{equation}
where the functions $f$ are certain logarithmically growing functions
defined in \cite{BarrZee, Chang}.

We calculate the electron EDM numerically as a function of
the LRSUSY model parameters $M_{B-L}$ and $M_R$ using
ROOT version 6.16 libraries. We let $M_{B-L}$ range
from $10^4$ to $10^{10}~\mathrm{GeV}$. For a given $M_{B-L}$, we let 
$M_R$ range from a low of $10^2 M_{B-L}$ to a high of 
$M_{B-L}^2 / M_{\mathrm{EW}}$. The lower bound on $M_R$ allows us to
safely break parity and left-right symmetry before breaking R-parity,
and the upper bound on $M_R$ allows the 
$\Omega$ fields in the left handed domain, which have a mass of about
$M_{B-L}^2 / M_R$, to stay heavier than the electroweak scale or the
supersymmetry breaking scale \cite{Aulakh:1997fq}. The lower bound on 
$M_{B-L}$ ensures that it is above any reasonable supersymmetry
breaking scale $M_S$ and so one can comfortably break the 
$U(1)_{B-L}$ gauge symmetry to reduce to the MSSM. In other words,
as argued in more detail in \cite{Aulakh:1997fq}, the low energy effective
theory of LRSUSY turns out to be the MSSM with strictly unbroken
R-parity, and so the lightest supersymmetric particle is stable.
The upper bound on $M_{B-L}$ follows from the consideration that
for $M_{B-L} \geq 10^{10}~\mathrm{GeV}$ we have to take 
$M_{R} \geq 10^{12}~\mathrm{GeV}$ which is rather high for parity
breaking.
The experimentally allowed region, where the electron EDM, is less than
$1.1 \times 10^{-29}~\mathrm{e~cm}$ \cite{ACME2}, is plotted as the
green hatched region in
the $(M_{B-L}, M_R)$-plane in Figure~\ref{fig:allowed}. 

\section{Leptogenesis constraints on LRSUSY}
The non-supersymmetric LR model was studied in the context of 
conventional electroweak baryogenesis mechanisms in 
\cite{MohapatraZhang, Frere}, and in a domain wall mediated 
baryogenesis via leptogenesis mechanism in \cite{ClineYajnik}.
The possibility of extending the latter mechanism to LRSUSY was 
indicated in \cite{Anjishnu}, but no concrete numerical calculations were 
performed there. In this paper, we address this deficiency.

Adequate amount of CP violation as well as strong loss of equilibrium
conditions have been major challenges for low energy baryogenesis. 
The presence of a moving domain wall, a topological defect, towards
the energetically disfavoured right handed domain
immediately guarantees a strong loss of equilibrium. The main LRSUSY model
per se does not explain why this happens, but we assume that this
occurs because of tiny effects like soft SUSY terms \cite{Anjishnu}
or Planck suppressed non-renormalisable terms \cite{Mishra} 
breaking exact left-right symmetry. A similar calculation has recently
been done \cite{SO10DomainWall} showing how Planck suppressed 
non-renormalisable terms 
can remove a pseudo domain wall in supersymmetric $\mathrm{SO}(10)$ GUT 
without conflicting with standard cosmology.
Given that the domain wall in LRSUSY has somehow disappeared
early enough so as not to conflict with present day cosmology,
we can exploit it to obtain leptogenesis and consequent
baryogenesis consistent with experimental bounds on baryon asymmetry.
This is done as follows.

The lepton-Higgs Yukawa part of the superpotential of LRSUSY is 
\cite{Aulakh:1997fq}
\begin{equation}
W_Y = 
\mathbf{h}_l^{(j)} L^T \tau_2 \Phi_j \tau_2 L_c +
i \mathbf{f} (L^T \tau_2 \Delta L + L_c^T \tau_2 \Delta_c L_c), 
\end{equation}
where $j = 1, 2$ and $\mathbf{h}$, $\mathbf{f}$ are $3 \times 3$
real symmetric matrices.
The Majorana mass terms above corresponding to the Yukawa coupling matrix
$\mathbf{f}$ are a source of lepton number violation. However, in 
LRSUSY they do not favour conventional thermal leptogenesis because at the
usual scale of thermal leptogenesis, the $B-L$ gauged symmetry is
unbroken \cite{Anjishnu}. That is why we have to resort to domain wall
mediated leptogenesis in LRSUSY. The lepton number violating decay of
the heavy Majorana RH neutrino will instead give rise to a lepton asymmetry
washout which will dilute any lepton asymmetry mediated by the domain
wall. Nevertheless, as we will see below, it is possible to obtain
baryon asymmetry consistent with experimental data for a certain region
of the LRSUSY parameter space.

Consider a domain wall moving slowly with speed $v_w$ in the $+x$
direction i.e. encroaching upon the energetically disfavoured RH domain.
The wall is assumed to stretch from $-L$ to $+L$ in the $x$ direction
and be flat in the $yz$ plane.
Slow speed means that $v_w < 1/\sqrt{3}$, the speed of sound in the
hot plasma, allowing one to get a solution to the chemical potential
of the neutrinos in terms of a fluid approximation \cite{Joyce}. 
Since the wall is assumed to move 
due to tiny energy
differences between the LH and RH domain, this is a reasonable assumption.
We consider the case of thick walls i.e. $2 L > 1 / T$, the de Broglie
wavelength of the neutrinos at temperature $T$ \cite{Joyce}. 
This is a reasonable
assumption ensuring that the mean free path of the neutrinos is smaller
than the wall thickness, leading to multiple interactions between
neutrinos and the
CP violating condensate in the wall and allowing a classical WKB
treatment of the neutrinos. A wall thickness of $10/T$ will be typical
in our analysis. We assume that the neutrino diffusion coefficient 
$D < 2L / (3v_w)$ \cite{Joyce}, allowing us to do a thermalised fluid 
analysis of the chemical potential of the LH neutrinos. The expression
for $D$ below easily satisfies this constraint.

Consider the interaction of LH neutrinos with the domain wall.
The LH neutrinos $\nu_L$ are massive in the RH domain due to the Majorana
Yukawa coupling to the $\SU(2)_L$-triplet Higgs field $\Delta$ which 
takes a vev in the RH domain. Conversely they are massless in the LH
domain where $\Delta$ has a vanishing vev. Due to the existence of
a spatially varying CP violating condenstate arising from the 
complex bidoubet vevs inside the wall, one
get an asymmetry between $\nu_L$ and its antiparticle $\bar{\nu}_L$ 
in terms of their 
reflection and transmission coefficients with respect to
the wall. There will be a preference for transmission of, say, $\nu_L$
from RH domain to LH domain through the wall. 
The LH neutrinos reflected back into RH domain from the wall quickly
equilibrate with their antiparticles around them because of the high
rate of helicity flipping of LH neutrinos owing to their large
Majorana mass in the RH domain. Thus the RH domain continues to have
almost no particle antiparticle asymmetry. In contrast, the transmitted
excess of LH neutrinos into LH domain survives because they are almost
massless in LH domain and hardly flip helicity. This leads to a 
particle antiparticle asymmetry in LH domain. Eventually as the
domain wall encroaches totally upon RH domain and destroys it, we end up
with an excess of leptons in our left handed Universe. Weak sphaeleron
processes convert a part of the early lepton excess into a baryon
excess till they go out of equilibrium as our
left handed Universe cools. Thus, one is left with a baryon excess
in the present day Universe.

The diffusion equation for the chemical potential $\mu$ of the 
LH neutrino in the wall rest frame is given by \cite{ClineYajnik}
\begin{equation}
\label{eq:Diffusion}
-D \mu'' + v_w \mu' + \Gamma \Theta(x) \mu = S(x),
\end{equation}
where $S(x)$ is the so-called source term defined below,
$D$ is the neutrino
diffusion coefficient, $v_w$ is the speed of the 
wall taken to be moving in the $+x$ direction, $\Theta(x)$ is the
step function which takes value one if $x$ is positive and zero
otherwise, $\Gamma$ is 
the rate of helicity flipping interactions at temperature $T$
which is very high in the RH domain due to the heavy
Majorana mass of the LH neutrino in the RH domain and zero in the
LH domain since the LH neutrino is almost massless in the LH domain.
Observe that the LH neutrino mass $m_\nu(x)$ is spatially dependent
and complex inside the wall since the neutrino couples to the spatially
varying complex bidoublet
Higgs and the triplet Higgs fields $\Delta$, $\Delta_c$ quickly
vanish inside the wall (see Figures~\ref{fig:Triplets-Delta}, \ref{fig:Phi-vevs}).

The source term, which is a CP-violating non-zero force if the 
neutrino mass $m_\nu(x)$ and the domain wall CP phase $\delta(x)$ are 
spatially varying, is given by \cite{ClineSUSY}
\begin{equation}
S(x) = 
-\frac{v_w D}{2 \Gamma}
\langle \frac{|p_x|}{E^2 \tilde{E}} \rangle
(m_\nu(x) \delta'(x))'',
\end{equation}
where $p_x$ is the $x$-component of the LH neutrino's momentum,
$E$ is the neutrino energy, $\tilde{E}$ is the related quantity
$\sqrt{m_\nu(x)^2 + p_x^2}$, and the angular brackets indicate thermal
averages. It was shown in 
\cite{ClineSUSY} that
\begin{equation}
\langle \frac{|p_x|}{E^2 \tilde{E}} \rangle =
\frac{e^{-a} - a E_1(a)}{2 T^2 a^2 K_2(a)},
\end{equation}
where $a = m_\nu(\infty) / T$, $E_1$ is the error function, $K_2$ is
the modified cylindrical Bessel function of the second kind and
the LH neutrino mass is evaluated deep inside the RH domain.
The source term is zero outside
the wall but non-zero inside.

The helicity flipping rate can be calculated by \cite{Joyce} 
\begin{equation}
\Gamma = \frac{\alpha_w^2 m_\nu(\infty)^2}{T},
\end{equation}
where $\alpha_w$ is the weak coupling constant evaluated at temperature 
$T$. The diffusion coefficient has the expression \cite{ClineSUSY},
\begin{equation}
D = \frac{\langle v_x^2 \rangle}{\Gamma}, 
~~~
\langle v_x^2 \rangle = 
\frac{3a+2}{a^2 + 3a + 2},
\end{equation}
where $v_x$ is the $x$-component of the neutrino velocity and $a$
is defined above.

We solve the diffusion equation (Equation~\ref{eq:Diffusion}) numerically
the using GSL version 2.6 library under the same settings as before,
and take the limiting value $\mu(-\infty)$
in the LH domain as the steady state chemical potential of the LH 
neutrino.
Then, the steady state neutrino-antineutrino asymmetry in the LH domain 
becomes \cite{ClineYajnik}
\begin{equation}
\Delta_{\nu_L} = \frac{T^2}{6} \mu(-\infty).
\end{equation}
To obtain the raw lepton asymmetry to entropy density ratio 
$\eta^{\mathrm{raw}}$, we need to divide
the above quantity by $\frac{2 \pi^2 g_* T^3}{45}$, where
$g_* \approx 110$ is the number of relativistic degrees of freedom.
Doing so gives us 
\begin{equation}
\eta^{\mathrm{raw}}_L = 
0.0035 \frac{\mu(-\infty)}{T}.
\end{equation}
Since the heavy neutrinos in this model have mass less than 
the temperature $M_{B-L}$, they can easily decay violating lepton
number. This process washes out most of the raw lepton asymmetry 
$\eta^{\mathrm{raw}}_L$. The surviving lepton asymmetry by entropy density
ratio becomes \cite{ClineYajnik}
\begin{equation}
\eta_L = 
\eta^{\mathrm{raw}} \cdot
10^{-4 \cdot 10^{-4} m_\nu M_{\mathrm{Planck}} v^{-2}} =
3.054 \cdot 10^{-11} \frac{\mu(-\infty)}{T},
\end{equation}
where $m_\nu$ is the mass of the 
heaviest light neutrino and $v = 174~\mathrm{GeV}$
is the SM Higgs vev. We take $m_\nu = 0.05~\mathrm{eV}$ 
from the Nu-FIT Group \cite{NuFit}. 

\begin{figure}[!hhh]
\begin{tikzpicture}
\node[anchor=south west,inner sep=0] (image) at (0,0) 
{\includegraphics[width=\textwidth]{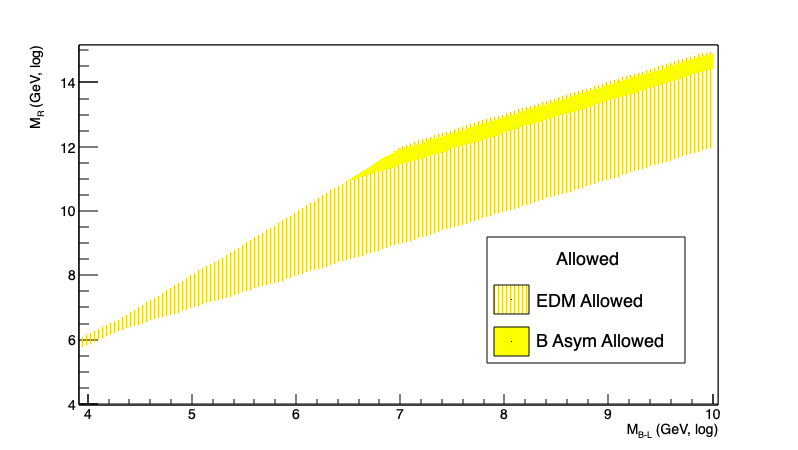}};
\begin{scope}[x={(image.south east)},y={(image.north west)}]
\draw[color=blue] (0.1, 0.245) -- (0.71, 0.90); 
\draw[color=blue] (0.35, 0.6) -- (0.35, 0.6) node [anchor=south]
{{\tiny $M_R = \frac{M_{B-L}^2}{M_{\mathrm{EW}}}$}}; 

\draw[color=blue] (0.1, 0.245) -- (0.89, 0.67); 
\draw[color=blue] (0.35, 0.35) -- (0.35, 0.35) node [anchor=north]
{{\tiny $M_R = 100 M_{B-L}$}}; 

\draw[color=blue] (0.80, 0.34) -- (0.80, 0.34) node [anchor=west]
{{\footnotesize $< 10^{-29}$ e cm}}; 
\draw[color=blue] (0.825, 0.25) -- (0.825, 0.25) node [anchor=west]
{{\footnotesize $10^{-11}$---$10^{-8}$}}; 
\end{scope}
\end{tikzpicture}
\caption{
The regions in the $(M_{B-L}, M_R)$-plane allowed by the
experimental bounds on baryon asymmetry in yellow and electron
EDM hatched green, plotted for the same parameter set as 
Fig.~\ref{fig:Triplets-Omega}-\ref{fig:Phi-phases}.
The upper blue line denotes the setting $M_{R} = M_{B_L}^2 / M_{EW}$ and
the lower blue line $M_R = 100 M_{B-L}$.
The baryon asymmetry allowed region is the narrow region entirely 
subsumed within that allowed
by the EDM constraint and overlaid on the latter. The two regions 
run almost parallel at upper boundary
with EDM allowed region somewhat larger. To be specific, 
$10^{6.5} < M_{B-L} < 10^{10}\,\mathrm{GeV}$
and bears out the philosophy of Eq. \eqref{eq:BLscaleseesaw} only 
towards the lowest end of the range.
}
\label{fig:allowed}
\end{figure}

Finally electroweak sphaelerons convert part of the lepton asymmetry
to baryon asymmetry starting from the temperature $M_{B-L}$ till
the universe cools to the  sphaeleron scale of about one TeV. This
gives the steady state baryon asymmetry to entropy density ratio
\cite{ClineYajnik}
\begin{equation}
\eta_B = \frac{28}{51} \eta_L = 
1.677 \cdot 10^{-11} \frac{\mu(-\infty)}{T}.
\end{equation}
The yellow strip in Figure~\ref{fig:allowed} shows the $(M_R, M_{B-L})$
tuples where the final baryon asymmetry to entropy density ratio is 
between $10^{-11}$ and $10^{-8}$, which can
thus provide a good explanation of the experimentally observed
baryon asymmetry of $6 \cdot 10^{-10}$ \cite{Canetti}. The parameter
space of the yellow strip is consistent with the experimental upper bound 
on electron EDM indicated by the green hatched region in  
Figure~\ref{fig:allowed}.

\section{Discussion and conclusion}
\label{sec:conclusion}
Armed with our numerical calculation tools,
we have made a strategic exploration of the parameter space of LRSUSY to find the
subregion consistent with the current experimental bounds on electron
EDM and baryon-antibaryon asymmetry.
We have varied the trilinear $\Omega \Phi \Phi$ coupling parameter 
$\alpha$ from $0.001$ to $0.1$, and the
mass parameters $\mu$ from $0.5$ to $1$ in order
to study the shape of the Higgs fields inside the wall and have
obtained similar results as described above. These ranges of the 
parameters were chosen
from the following considerations. Values of $\alpha$ greater than $0.1$ 
make the soft terms not so soft anymore, and values of mass parameters
$\mu$ outside the above range either make the bidoublets heavier than
the largest mass scale $M_R$ or make our fine tuning
ineffective. The manifestation of the limiting values of the parameters
shows up in the observation that for $\mu_{11}$, $\mu_{22}$ smaller
than $0.5 M_R$, or for $\alpha > 0.1$, there is no value of 
$M_{B-L}$ between $10^4$ to $10^{10}\,\mathrm{GeV}$ consistent with
both electron EDM and baryon asymmetry experiments.

\begin{figure}[!hhh]
\includegraphics[width=0.9\textwidth]{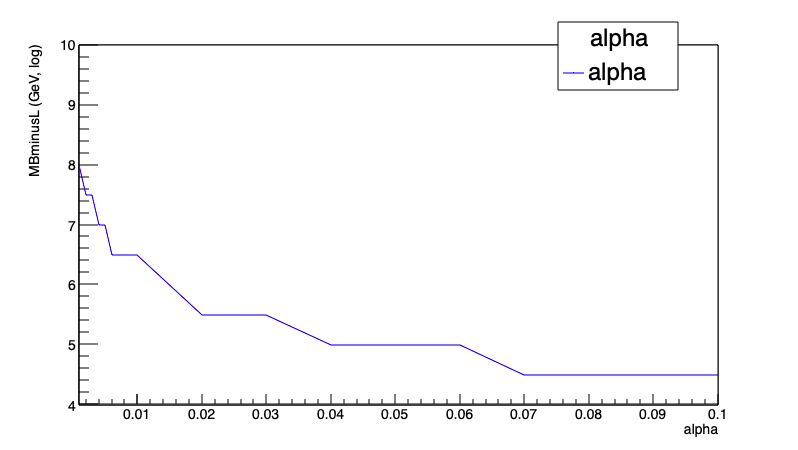}
\caption{The lowest value of $M_{B-L}$ consistent with both electron
EDM and baryon asymmetry experiments plotted as a function of $\alpha$.
}
\label{fig:alphaMBminusL}
\end{figure}

For parameter values in the above ranges, we see that the lowest
$M_{B-L}$ consistent with the two experiments is $10^{4.5}\,\mathrm{GeV}$.
The lowest allowed value of $M_{B-L}$ is most sensitive to $\alpha$,
and is a decreasing function of $\alpha$. Figure~\ref{fig:alphaMBminusL}
shows its variation when $\alpha$ ranges from $0.001$ to $0.1$.

It turns out that the lowest allowed value of $M_{B-L}$ is hardly 
sensitive to the mass parameters and variations in wall thickness
and wall velocity. Thus, by scanning the parameter space we can
conclude that for consistency with both electron EDM and baryon
asymmetry experiments, the scale $M_{B-L}$ of $B-L$-symmetry breaking 
must be larger than $10^{4.5}\,\mathrm{GeV}$, and over the whole allowed range
must be significantly less that $M_{B-L}^2 / M_{EW}$.
These novel bounds provide the most stringent constraints on 
the parameter space of LRSUSY by far.

More interestingly, our implication that $M_R$ must be significantly
less than $M_{B-L}^2 / M_{EW}$ accords with the simplifying proposal  of
Aulakh et al. \cite{Aulakh:1997fq} viz. $M_R \gtrsim M_{B-L}^2 / M_{EW}$, as discussed
at \eqref{eq:BLscaleseesaw}, only at the lowest allowed value of $M_{B-L}$.
Aulakh et al.'s proposal arose from gravity mediated TeV scale SUSY breaking where the gravitino is 
not much heavier than $M_{EW}$. Within the validity of this proposal this analysis makes a specific prediction
of the $M_{B-L}$ and hence the $M_R$.  Since LHC has not discovered any signatures of SUSY,
TeV scale SUSY breaking is now a highly unlikely possibility though an exciting one to confirm if true.

On the other hand we can reject the $R$ charge proposed 
in \cite{Aulakh:1997fq}, in order to allow
the fact that the scale of parity breaking and $(B-L)$-symmetry 
breaking are not maximally far apart.
The parameter $m_\Omega$ could then be intrinsic to the superpotential. 
In this case our results are perfectly consistent with PeV scale supersymmetry \cite{Wells}
also considered in the recent work \cite{Dhuria}, where the
gravitino can be much heavier.  This places LRSUSY outside the experimental reach of colliders in 
the near future.  On the other hand, the discovery of a non-zero electron EDM value can be taken to 
be a hint to a narrow range for the $M_R$ scale assuming a world with a renormalisable supersymmetric 
left-right model. Finally it is interesting that two very different low-energy probes viz. baryon asymmetry 
and electron EDM  provide a strong constraint on the allowed parameter space of LRSUSY.

\bibliography{cpviolation}

\appendix

\section{Euler-Lagrange equations for spatially varying Higgs vevs}
\label{sec:EL}
The Euler-Lagrange equations are
written explicitly below, where $\ddot{f}$ represents the double derivative
of vev of field $f$ with respect to $x$ and $F_{(\Phi_i)_j}$ denotes
the $(j,j)$ matrix element of the vev $\vev{F_{\Phi_i}}$. The 
expressions for the vevs of the F-terms and D-terms can be found
in Equations~\ref{eq:FTermVevs} and \ref{eq:DTermVevs} respectively.
There are eight bidoublet field vevs $r_1, i_1, \ldots, r_2', i_2'$ which
are the real and imaginary parts of the vev $k_1, \ldots, k_2'$
respectively. There are 4 triplet fields vevs viz. $d$, $\bar{d}$,
$d_c$, $\bar{d}_c$. All these vevs vary as a function of distance $x$
from the centre of the wall, but we write field $f$ in the equations
below and not $f(x)$ for clarity of notation. The vevs of the
fields $\omega(x)$, $\omega_c(x)$ below take their values from
the ansatz in Equation~\eqref{eq:OmegaAnsatz}.
\begin{equation}
\label{eq:EL}
\begin{array}{r c l}
\ddot{r}_1 
& = &
4 \alpha \mathrm{Re}(F_{\Omega} + F_{\Omega_c}) r_2' +
8 \alpha \mathrm{Im}(F_{\Omega}) i_2' +
8 \mu_{11} \mathrm{Re}(F_{(\Phi_1)_2}) +
8 (\mu_{12} + \alpha(\omega - \omega_c)) \mathrm{Re}(F_{(\Phi_2)_2}) \\
&  &
{} +
8 g^2 r_1 D_{L,3} - 2 \mu_1^2 r_1  -
4 \mu_3^2 (r_1' \cos \beta_3 + i_1' \sin \beta_3) -
2 \mu_5^2 (r_2' \cos \beta_5 + i_2' \sin \beta_5), \\
\ddot{i}_1 
& = &
-4 \alpha \mathrm{Re}(F_{\Omega} + F_{\Omega_c}) i_2' +
8 \alpha \mathrm{Im}(F_{\Omega}) r_2' +
8 \mu_{11} \mathrm{Im}(F_{(\Phi_1)_2}) +
8 (\mu_{12} + \alpha(\omega - \omega_c)) \mathrm{Im}(F_{(\Phi_2)_2}) \\
&  &
{} +
8 g^2 i_1 D_{L,3} - 2 \mu_1^2 i_1  -
4 \mu_3^2 (-i_1' \cos \beta_3 + r_1' \sin \beta_3) -
2 \mu_5^2 (-i_2' \cos \beta_5 + r_2' \sin \beta_5), \\
\ddot{r}_2 
& = &
-4 \alpha \mathrm{Re}(F_{\Omega} + F_{\Omega_c}) r_1' -
8 \alpha \mathrm{Im}(F_{\Omega}) i_1' +
8 (\mu_{12} - \alpha(\omega - \omega_c)) \mathrm{Re}(F_{(\Phi_1)_2}) +
8 \mu_{22} \mathrm{Re}(F_{(\Phi_2)_2}) \\
&  &
{} +
8 g^2 r_2 D_{L,3} - 2 \mu_2^2 r_2  -
4 \mu_4^2 (r_2' \cos \beta_4 + i_2' \sin \beta_4) -
2 \mu_5^2 (r_1' \cos \beta_5 + i_1' \sin \beta_5), \\
\ddot{i}_2 
& = &
4 \alpha \mathrm{Re}(F_{\Omega} + F_{\Omega_c}) i_1' -
8 \alpha \mathrm{Im}(F_{\Omega}) r_1' +
8 (\mu_{12} - \alpha(\omega - \omega_c)) \mathrm{Im}(F_{(\Phi_1)_2}) +
8 \mu_{22} \mathrm{Im}(F_{(\Phi_2)_2}) \\
&  &
{} +
8 g^2 i_2 D_{L,3} - 2 \mu_2^2 i_2  -
4 \mu_4^2 (-i_2' \cos \beta_4 + r_2' \sin \beta_4) -
2 \mu_5^2 (-i_1' \cos \beta_5 + r_1' \sin \beta_5), \\
\ddot{r}'_1 
& = &
-4 \alpha \mathrm{Re}(F_{\Omega} + F_{\Omega_c}) r_2 -
8 \alpha \mathrm{Im}(F_{\Omega}) i_2 +
8 \mu_{11} \mathrm{Re}(F_{(\Phi_1)_1}) +
8 (\mu_{12} - \alpha(\omega - \omega_c)) \mathrm{Re}(F_{(\Phi_2)_1}) \\
&  &
{} -
8 g^2 r_1' D_{L,3} - 2 \mu_1^2 r_1'  -
4 \mu_3^2 (r_1 \cos \beta_3 + i_1 \sin \beta_3) -
2 \mu_5^2 (r_2 \cos \beta_5 + i_2 \sin \beta_5), \\
\ddot{i}'_1 
& = &
4 \alpha \mathrm{Re}(F_{\Omega} + F_{\Omega_c}) i_2 -
8 \alpha \mathrm{Im}(F_{\Omega}) r_2 +
8 \mu_{11} \mathrm{Im}(F_{(\Phi_1)_1}) +
8 (\mu_{12} - \alpha(\omega - \omega_c)) \mathrm{Im}(F_{(\Phi_2)_1}) \\
&  &
{} -
8 g^2 i_1' D_{L,3} - 2 \mu_1^2 i_1'  -
4 \mu_3^2 (-i_1 \cos \beta_3 + r_1 \sin \beta_3) -
2 \mu_5^2 (-i_2 \cos \beta_5 + r_2 \sin \beta_5), \\
\ddot{r}'_2 
& = &
4 \alpha \mathrm{Re}(F_{\Omega} + F_{\Omega_c}) r_1 +
8 \alpha \mathrm{Im}(F_{\Omega}) i_1 +
8 (\mu_{12} + \alpha(\omega - \omega_c)) \mathrm{Re}(F_{(\Phi_1)_1}) +
8 \mu_{22} \mathrm{Re}(F_{(\Phi_2)_1}) \\
&  &
{} -
8 g^2 r_2' D_{L,3} - 2 \mu_2^2 r_2'  -
4 \mu_4^2 (r_2 \cos \beta_4 + i_2 \sin \beta_4) -
2 \mu_5^2 (r_1 \cos \beta_5 + i_1 \sin \beta_5), \\
\ddot{i}'_2 
& = &
-4 \alpha \mathrm{Re}(F_{\Omega} + F_{\Omega_c}) i_1 +
8 \alpha \mathrm{Im}(F_{\Omega}) r_1 +
8 (\mu_{12} + \alpha(\omega - \omega_c)) \mathrm{Im}(F_{(\Phi_1)_1}) +
8 \mu_{22} \mathrm{Im}(F_{(\Phi_2)_1}) \\
&  &
{} -
8 g^2 i_2' D_{L,3} - 2 \mu_2^2 i_2'  -
4 \mu_4^2 (-i_2 \cos \beta_4 + r_2 \sin \beta_4) -
2 \mu_5^2 (-i_1 \cos \beta_5 + r_1 \sin \beta_5). 
\end{array}
\end{equation}
\begin{equation}
\begin{array}{rcl}
\ddot{d}
& = &
2 a \mathrm{Re}(F_{\Omega}) \bar{d} +
2 F_{\bar{\Delta}} (m_\Delta + a \omega), \\
\ddot{\bar{d}}
& = &
2 a \mathrm{Re}(F_{\Omega}) d +
2 F_{\Delta} (m_\Delta + a \omega), \\
\ddot{d}_c
& = &
2 a \mathrm{Re}(F_{\Omega_c}) \bar{d}_c +
2 F_{\bar{\Delta_c}} (m_\Delta + a \omega_c), \\
\ddot{\bar{d}}_c
& = &
2 a \mathrm{Re}(F_{\Omega_c}) d_c +
2 F_{\Delta_c} (m_\Delta + a \omega_c).
\end{array}
\end{equation}

\section{Temperature dependent mass matrix of neutral bidoublet Higgs}
\label{sec:tempmass}
Here we display the temperature dependent mass matrix corresponding to 
Equations~\ref{eq:VSUSY}, \ref{eq:Vsoft}.
This is a symmetric $8 \times 8$ matrix in the fields
$r_1$, $i_1$, $r_2$, $i_2$, $r'_1$, $i'_1$, $r'_2$, $i'_2$. The
mass parameters $\mu$ are assumed to be corrected for the
temperature of $T = M_{B-l}$.
Following \cite{Cline2HDM}, we can write the matrix with the 
zero temperature Debye mass corrections as shown below. In obtaining 
the domain wall profiles we use $T = 0$ and for the electron EDM
calculation we use $T = M_{B-L}$.
\begin{equation}
\begin{array}{rcl}
M_{r_1,r_1}
& = &
M_{i_1,i_1} =
\mu_{11}^2 + (\mu_{12} - \alpha M_R)^2 - \mu_1^2
- \frac{g^2 T^2}{6} 
\\
M_{r_2,r_2}
& = &
M_{i_2,i_2} = 
(\mu_{12} + \alpha M_R)^2 + \mu_{22}^2 - \mu_2^2
- \frac{g^2 T^2}{6} 
\\
M_{r'_1,r'_1}
& = &
M_{i'_1,i'_1} =
\mu_{11}^2 + (\mu_{12} + \alpha M_R)^2 - \mu_1^2
- \frac{g^2 T^2}{6} 
\\
M_{r'_2,r'_2} 
& = &
M_{i'_2,i'_2} =
(\mu_{12} - \alpha M_R)^2 + \mu_{22}^2 - \mu_2^2
- \frac{g^2 T^2}{6} 
\\
& & \\
M_{r_1,i_1} 
& = &
M_{r_2,i_2} =
M_{r'_1,i'_1} =
M_{r'_2,i'_2} =
M_{r_1,i_2} =
M_{i_1,r_2} =
M_{r'_1,i'_2} =
M_{i'_1,r'_2} =
0
\\
& & \\
M_{r_1,r_2}
& = &
\mu_{11}(\mu_{12} + \alpha M_R) + (\mu_{12} - \alpha M_R)\mu_{22}
\\
M_{i_1,i_2}
& = &
\mu_{11}(\mu_{12} + \alpha M_R) + (\mu_{12} - \alpha M_R)\mu_{22}
\\
M_{r'_1,r'_2}
& = &
\mu_{11}(\mu_{12} - \alpha M_R) + (\mu_{12} + \alpha M_R)\mu_{22}
\\
M_{i'_1,i'_2}
& = &
\mu_{11}(\mu_{12} - \alpha M_R) + (\mu_{12} + \alpha M_R)\mu_{22}
\\
& & \\
M_{r_1,r'_1}
& = &
-M_{i_1,i'_1} = 
-2\mu_3^2 \cos \beta_3
\\
M_{r_1,i'_1}
& = &
-M_{i_1,r'_1} =
-2\mu_3^2 \sin \beta_3
\\
& & \\
M_{r_1,r'_2}
& = &
-M_{i_1,i'_2} =
M_{r_2,r'_1} =
-M_{i_2,i'_1} =
-\mu_5^2 \cos \beta_5 
\\
M_{r_1,i'_2}
& = &
M_{i_1,r'_2} =
M_{r_2,i'_1} =
M_{i_2,r'_1} =
-\mu_5^2 \sin \beta_5
\\
& & \\
M_{r_2,r'_2}
& = &
-M_{i_2,i'_2} =
-2\mu_4^2 \cos \beta_4
\\
M_{r_2,i'_2}
& = &
-M_{i_2,r'_2} =
-2\mu_4^2 \sin \beta_4
\end{array}
\end{equation}

\end{document}